\documentclass[twocolumn,showpacs,amsmath,aps,amssymb]{revtex4-1}
\usepackage{graphicx}
\usepackage{dcolumn}
\usepackage{bm}
\usepackage{color}

\begin{document}

\title{Deformed Phase Space Kaluza-Klein cosmology and late time acceleration.}
\author{M. Sabido}
\email{msabido@fisica.ugto.mx}
\affiliation{ Departamento  de F\'{\i}sica de la Universidad de Guanajuato,\\
 A.P. E-143, C.P. 37150, Le\'on, Guanajuato, M\'exico
 }%

\author{C. Yee-Romero}
 \email{carlos.yee@uabc.edu.mx}
\affiliation{  Departamento de Matem\'aticas, Facultad de Ciencias\\
 Universidad Aut\'onoma de Baja California, Ensenada, Baja California, M\'exico\\}%

\begin{abstract}
The effects of phase space deformations on Kalutza-Klein cosmology  are studied. The deformation is introduced by modifying the symplectic structure of the minisuperspace variables. In the deformed model,  we find an accelerating scale factor and  therefore infer the existence of an effective cosmological constant from the phase space deformation parameter $\beta$. 
\end{abstract}
 \maketitle
 \newpage

\section{Introduction}
Current cosmological observations are best described by $\Lambda CDM$ model \cite{planck}, where
the current  acceleration of the universe is attributed to   $\Lambda$. Unfortunately there  are several theoretical  problems  connected to the cosmological constant, making  the cosmological constant one of the central issues of modern day  physics. There is a belief that the solution will come from an unconventional approach in fundamental physics (i.e. arguments are given that  UV/IR mixing mechanism is needed \cite{polchinski}), suggesting the need of new physics. 
One approach to study the cosmological constant, lies in noncommutative space-time,  from which  several approaches to noncommutative gravity  where proposed \cite{GarciaCompean:2003pm}. All of these formulations showed that the end result of a noncommutative theory of gravity, is a highly nonlinear theory. In order to study the effects of noncommutativity, noncommutative cosmology was presented in  \cite {GarciaCompean:2001wy}.  
Although the deformations of the minisuperspace where originally studied at the quantum level,  classical noncommutative formulations have been proposed \cite{Barbosa:2004kp}. The  idea  is based on the assumption that modifying the Poisson brackets of the classical theory gives the noncommutative equations of motion. A more general deformation of the Poisson algebra of the minisuperspace variables gives rise to deformed phase space cosmology. 
Phase space deformations give rise to two generally different interpretations known as  the {\it ``C-frame''} and the {\it ``NC-frame''}, that in general are not physically equivalent \cite{brasil}. For this reason we must be careful when reaching physical conclusions in phase space deformations. 

These ideas have been applied in the context of the late time acceleration of the universe. In \cite{vakili,tachy} the authors study the late time effects of    minisuperspace deformations on cosmology, suggesting a relationship between late time acceleration  and the deformation parameters.
This was not the first time evidence was found on the possible effects of 
phase space deformations in the cosmological scenario. In \cite{kallosh} it is argued that there is a possible relation between the 4D cosmological constant and the noncommutative parameter of the compactified space in string theory.  A more direct  connection with the cosmological constant problem has been  addressed in \cite{nclambda}, where it is shown that by means of minisuperspace noncommutativity a small cosmological constant arises, and seems to alleviate the discrepancy between the calculated and observed vacuum energy density.

In this paper, we study deformed phase space Kalutza-Klein (KK) cosmology. We introduce the phase space deformation in the minisuperspace variables, and is achieved by modifying the symplectic structure. Finally  we derive the effective cosmological constant  that depends on the deformation parameters $\theta$ and $\beta$. 
The paper is organised as follows, in section II, we start with an  empty  (4+1) dimensional Kaluza-Klein universe with cosmological constant and an FRW metric. In Section III we introduce  the deformation in the phase space constructed from the minisuperspace variables and their conjugate momenta. Section IV is devoted for  conclusions and final remarks.

\section{The  Model}
We start with  an empty $(4+1)$ theory of gravity with cosmological constant $\Lambda$ as shown in \cite{darabi}. The action takes the form
\begin{equation}
I=\int\sqrt{-g}\left(R-\Lambda\right)dtd^3rd\rho.\label{he}
\end{equation} 
We are interested in cosmology, so an FRW type metric is assumed
\begin{equation}
ds^2=-dt^2+\frac{a^2(t)dr^idr^i}{\left(1+\frac{\kappa r^2}{4}\right)^2}+\phi^2(t)d\rho^2,
\end{equation}  
where $\kappa=0,\pm1$ and $a(t)$, $\phi(t)$ are the scale factors of the universe and the compact dimension. Substituting this metric in Eq. (\ref{he}), we obtain an effective lagrangian that only depends on $(a,\phi)$
\begin{equation}
L=\frac{1}{2}\left(a\phi\dot{a}^2+a^2\dot{a}\phi-\kappa a\phi+\frac{1}{3}\Lambda a^3\phi\right).\label{L1}
\end{equation}
Using the variables  
\begin{equation}
x=\frac{1}{\sqrt{8}}\left(a^2+a\phi-\frac{3\kappa}{\Lambda}\right),~y=\frac{1}{\sqrt{8}}\left(a^2-a\phi-\frac{3\kappa}{\Lambda}\right),\label{3}
\end{equation}
the Hamiltonian for the model is   
\begin{equation}
H=\frac{1}{2}\left[\left(P_{x}^2+\omega^2 x^2\right)-\left(P_{y}^2+\omega^2 y^2\right)\right],\label{HC}
\end{equation}
with $\omega^2=-\frac{2\Lambda}{3}$. This Hamiltonian describes an isotropic oscillator-ghost-oscillator system 
and is a first-class constraint,  as is usual in general relativity. Since we do not have second class constraints in the model we will continue to work with the usual   Poisson brackets and the commutation between the phase space variables 
\begin{equation}
\{ x_i, x_j \} = 0, \; \{ P_{x_i} , P_{y_j} \} = 0,\;\{x_i, P_{x_j} \}  = \delta_{ij}. \label{classicPoisson}
\end{equation}
 
The quantum model is obtained by following the canonical formalism, from (\ref{HC}) we can construct the Wheeler-DeWitt (WDW) equation, and get  the corresponding quantum cosmology for the model at hand. This is achieved by making the usual identifications $p_x=-i\partial/\partial{x}$ and $p_y=-i\partial/\partial{y}$,
\begin{equation}
\left[\frac{\partial^2}{\partial x^2}-\frac{\partial^2}{\partial y^2}+\omega^2(x^2 -y^2)\right]\Psi(x,y)=0.\label{ham}
\end{equation}
This equation gives the quantum description of the  model and the information about the quantum behaviour would be encoded in the wave function $\Psi(x,y)$.

\section{Deformed Space Model}

Canonical quantum cosmology, is described by the WDW equation where the quantization is performed on the minisuperspace variables. An alternative   to study quantum mechanical effects, is to introduce deformations to the phase space of the system, this approach can be considered part of deformation quantization \cite{hugo}.  The original ideas of a deformed minisuperspace, where done in connection with noncommutative cosmology \cite{GarciaCompean:2001wy}, by introducing a deformation to the minisuperspace in order to incorporate an effective noncommutativity. Therefore studying cosmological models in deformed phase could be interpreted as studying quantum effects to cosmological solutions \cite{vakili}.
In the deformed phase space approach, the deformation is introduced by the Moyal brackets $\{f,g\}_{\alpha}=f\star_{\alpha}g-g\star_{\alpha}f$, were the product between functions is replaced by  the Moyal product
\[
(f\star{g})(x)=\exp{\left[\frac{1}{2}\alpha^{ab}\partial_{a}^{(1)}\partial_{b}^{(2)}\right]}f(x_{1})g(x_2)\vert_{x_1=x_2=x}.\nonumber
\]
The resulting  $\alpha$ deformed algebra for the phase space variables is
\[
\{x_i,x_j\}_{\alpha}=\theta_{ij}, \;\{x_i,P_j\}_{\alpha}=\delta_{ij}+\sigma_{ij},\; \{P_i,P_j\}_{\alpha}=\beta_{ij}.
\] 
We can construct this algebra, with ``different" Poisson brackets. For the first case,  the brackets are the $\alpha$ deformed ones and are related to the  Moyal product. For the deformed phase space model  the brackets are the usual Poisson brackets. To construct a deformed poisson algebra 
we will follow the approach in \cite{vakili,tachy}.

We start with
 the following transformation on the classical phase space variables $\{x,y,P_x,P_y\}$, that satisfy the usual Poisson algebra
\begin{eqnarray}
\widehat{x}=x+\frac{\theta}{2}P_{y}, \qquad \widehat{y}=y-\frac{\theta}{2}P_{x},\nonumber \\
\widehat{P}_{x}=P_{x}-\frac{\beta}{2}y, \qquad \widehat{P}_{y}=P_{y}+\frac{\beta}{2}x.\label{nctrans}
\end{eqnarray}
These new variables satisfy a deformed algebra
\begin{equation}
\{\widehat{y},\widehat{x}\}=\theta,\; \{\widehat{x},\widehat{P}_{x}\}=\{\widehat{y},\widehat{P}_{y}\}=1+\sigma,\; \{\widehat{P}_y,\widehat{P}_x\}=\beta,\label{dpa}
\end{equation}
where $\sigma=\theta\beta/4$. Furthermore, as in \cite{vakili, tachy}, we assume that the deformed variables satisfy the same relations as their commutative counterpart.

Now that we construct the deformed theory, first we start with a Hamiltonian which is formally analogous to Eq.(\ref{HC}) but constructed with the variables that obey the modified algebra Eq.(\ref{dpa})
\begin{eqnarray}\label{HNC}
 H&=& \left( \frac{1}{2} \widehat{P}_x^2 + \frac{{\omega}^2}{2} \widehat{x}^2\right) - \left( \frac{1}{2} \widehat{P}_y^2 + \frac{{\omega}^2}{2} \widehat{y}^2 \right)\\
&=& \frac{1}{2}\left[\left( P_{x}^2-P_{y}^2\right )-\gamma^2(x P_{y}+ yP_{x}) + \widetilde{\omega}^2(x^2- y^2)\right], \nonumber
\end{eqnarray}
where we have used the change of variables Eq.(\ref{nctrans}) and  the following definitions
\begin{equation} 
\widetilde{\omega}^2=\frac{\omega^2-\frac{\beta^2}{4}}{1-\frac{\omega^2\theta^2}{4}}, \quad \gamma^2=\frac{\beta-\omega^2\theta}{1-\frac{\omega^2\theta^2}{4}}.\label{omegas}
\end{equation}
The WDW equation is obtained by the usual prescription on the deformed Hamiltonian Eq.(\ref{HNC}). The meaning of the first term in Eq.(\ref{omegas}) is straightforward, from the definition of $\omega$  the cosmological constant is related to the oscillator frequency, then modifications to the oscillator frequency will imply modifications to the effective cosmological constant.
Then   $\widetilde{\omega}$ gives the effective cosmological constant $\tilde\Lambda_{eff}$ in the context of the WDW equation \cite{eri}.
The case $\beta=0$, is equivalent to the standard noncommutative minisuperspace model that was presented in \cite{darabi} and used as a solution to the Hierarchy problem. 
 For the physical meaning of $\gamma$ we rewrite  Eq.(\ref{HNC})  as the original Hamiltonian Eq.(\ref{HC}) in the presence of a constant  ``magnetic field" $B$.  By using the vector potential $\vec{A}=(\frac{\gamma^2}{2}{y},-\frac{\gamma^2}{2}{x})$, we find that 
$B=-\gamma^2$.

It is important to remember that in noncommutative cosmology two different physical theories arise \cite{brasil}. One that considers the variables $x$ and $y$ as fundamental and another based on $\widehat{x}$ and $\widehat{y}$. The first theory, is interpreted as a ``commutative'' theory with a modified interaction, this theory is referred as being realised in the ``{\it C-frame}". 
The second theory which  privileges the variables $\widehat{x}$ and $\widehat{y}$, is a theory with ``noncommutative" variables but with the standard interaction and is referred to as realised in the ``{\it NC-frame}". One usually privileges one of the frames or assumes that the differences between them are negligible, but in some cases there can be dramatic differences in the physics on each frame. Although the physical implications between the frames can be startlingly different \cite{brasil}, there are cases where  the predictions are very similar. In those instances, one can take advantage of the formulation in the ``{\it C-frame}" and have very clear interpretation of the deformation. The appearance of these frames is also present in deformed phase space cosmology  and therefore the two frames must be studied. In particular, for the model under study, we have a very simple interpretati—n. In the ``{\it NC-frame}'' is the ghost oscillator in deformed phase space,  in the ``{\it C-frame}'' is the usual  commutative ghost oscillator in the presence of a ``magnetic field" $B$.

To obtain the dynamics for the model, we derive the equations of motion from the Hamiltonian Eq.(\ref{HNC})
\begin{eqnarray}
\dot{x}&=& P_{x}-\frac{1}{2}\gamma^2{y},\qquad {\dot{P}_{x}}=\frac{1}{2}\gamma^2 P_{y}-\widetilde{\omega}^2{x}, \nonumber \\
\dot{y}&=& -P_{y}-\frac{1}{2}\gamma^2{x},\quad  {\dot{P}_{y}}=\frac{1}{2}\gamma^2 P_{x}+\widetilde{\omega}^2{y}, \label{nceqm}
\end{eqnarray}
and get
The solutions for the variables ${x}(t)$ and ${y}(t)$ in the ``{\it C-frame}" are
\begin{equation}\nonumber
{x}(t)=\eta_0\:e^{\frac{-\gamma^2}{2}t} \cosh\left(\omega^{\prime}t+\delta_1\right)-\zeta_0\:e^{\frac{\gamma^2}{2}t} \cosh\left(\omega^{\prime}t+\delta_2\right),\label{xnc}
\end{equation}
\begin{equation}
{y}(t)=\eta_0\:e^{\frac{-\gamma^2}{2}t} \cosh\left(\omega^{\prime }t +\delta_1\right)+ \zeta_0\:e^{\frac{\gamma^2}{2}t} \cosh\left(\omega^{\prime }t+\delta_2 \right),\label{ync}
\end{equation}
where $\omega^{\prime 2 }=-\widetilde{\omega}^2$. For $\omega^{'2}<0$, the hyperbolic functions are replaced by harmonic functions.
There is a different solution for $\beta=2\omega$,  in the ``{\it C-frame}" 
\begin{eqnarray}
x(t)&=& (a+bt) e^{\frac{-\gamma^2}{2}t}+ (c+dt) e^{\frac{\gamma^2}{2}t},\\
y(t)&=& (a+bt) e^{\frac{-\gamma^2}{2}t}- (c+dt) e^{\frac{\gamma^2}{2}t}.\nonumber
\end{eqnarray}
To compute the volume of the universe in the ``{\it C-frame}" we use Eq.(\ref{3}) and Eq.(\ref{ync}). For $\omega'^2>0$ we get
\begin{equation}
a^3(t)=\left (\frac{3 k}{\Lambda}+V_0 e^{-\frac{\gamma^2}{2}t}\cosh{\left(\omega' t\right)}\right)^\frac{3}{2}\label{eq18},
\end{equation}
where we have taken $\delta_1=\delta_2=0$. For the case $\omega^{'2}<0$, the hyperbolic function is replaced by a harmonic function. For the case $\beta=2\omega$, the volume  is given by
\begin{equation}
a^3(t)=\left (\frac{3 k}{\Lambda}+V_0(1+V_1t) e^{-\frac{\gamma^2}{2}t}\right)^\frac{3}{2}\label{eq19},
\end{equation}
$V_0$, and  $V_1$ are constructed from the integration constants.

To find the dynamics  in the ``{\it NC-frame}" we start from the ``{\it C-frame}" solutions and use Eq.(\ref{nctrans}), we get 
\begin{equation} 
\widehat{a}^3(t) = \left\{ \begin{array}{ll}
        \left [\frac{3 k}{\Lambda}+ \widehat{V}_{0} e^{-\frac{\gamma^2}{2}t}\left( \cosh{\left(\omega' t\right)}-  \frac{\omega' \theta}{2}\sinh{\left(\omega' t\right)}\right) \right]^\frac{3}{2}\\ \hspace{5.3cm} \textup{for }\omega'>0;\\
        \left (\frac{3 k}{\Lambda}+\widehat{V}_{0}(1+\frac{\theta\widehat{V}_{1}}{2}+\widehat{V}_{1}t) e^{-\frac{\gamma^2}{2}t}\right)^\frac{3}{2}\\ \hspace{5.3cm} \textup{for }\omega'=0;\\
        \left [\frac{3 k}{\Lambda}+ \widehat{V}_{0} e^{-\frac{\gamma^2}{2}t}\left( \cos{\left(|\omega'| t\right)}-  \frac{|\omega'| \theta}{2}\sin{\left(|\omega'| t\right)}\right) \right]^\frac{3}{2}\\ \hspace{5.3cm} \textup{for }\omega'<0;\\
        \end{array} \right.\label{volume_nc}
\end{equation}
where $\widehat{V}_{0}$ is the initial volume in the {\it ``NC-frame"}. We can see that for $\theta=0$ the description in the two frames is the same. 

We are interested in the evolution of the model. Following \cite{vakili,tachy}, we will focus our discussion in the late time behaviour of the deformed phase space cosmology. Let us start by analysing the asymptotic behaviour of the volume for $t>>1$ in the two frames.
In both frames we see that only  for $\omega'^2\geq0$ the volume is positive  and therefore only consider that case as physically relevant.

Taking $\omega'^2=0$, fixes the value of $\beta=2\omega$.  For this case the asymptotic behaviour for the volume (in both frames) is $V(t)\sim Exp(-\gamma^2 t)$ with $\gamma^2=\frac{4\beta}{4+\beta\theta}$, therefore is inconsistent with the current acceleration of our universe. 

The only remaining possibility is $\omega'^2>0$, to analyze this case we will consider the following:
 $\Lambda_{eff}>0$, this in order to have an accelerating universe for $t>>1$ and
 $a^3(t)>0 $ to have a physical solution,
this will impose restrictions to the deformation parameters.

Following  \cite{tachy},  to find an expression for $\Lambda_{eff}$, we compare the de-Sitter cosmology scale factor with the deformed phase space model scale factor in the limit $t\to \infty$. In this limit, the first term in Eq.(\ref{volume_nc})  has an exponential form and behaves as the de-Sitter volume. Remembering the definition $\omega^2=-\frac{2}{3}\Lambda$ we get the following expression for the de-Sitter cosmological constant $\Lambda_{eff}$ as a function of  the noncommutative parameters
\begin{equation}
\Lambda_{eff} =\frac{3}{16}\left(\sqrt{\frac{\beta^2+\frac{8}{3}\Lambda}{1+\frac{1}{6}\Lambda\theta^2}}- \frac{\beta+\frac{2}{3}\Lambda\theta}{1+\frac{1}{6}\Lambda\theta^2}\right )^2\label{lambda}.
  \end{equation}
 The same result is obtained from Eq.(\ref{eq18}), therefore this expression is valid in the two frames.
 Imposing the requirement $\Lambda_{eff}>0$,  for $\Lambda>0$ we find a simple  condition over the deformation parameters $4-\beta\theta\ne0$.

\begin{figure}
\begin{center}
\includegraphics[width=8cm]{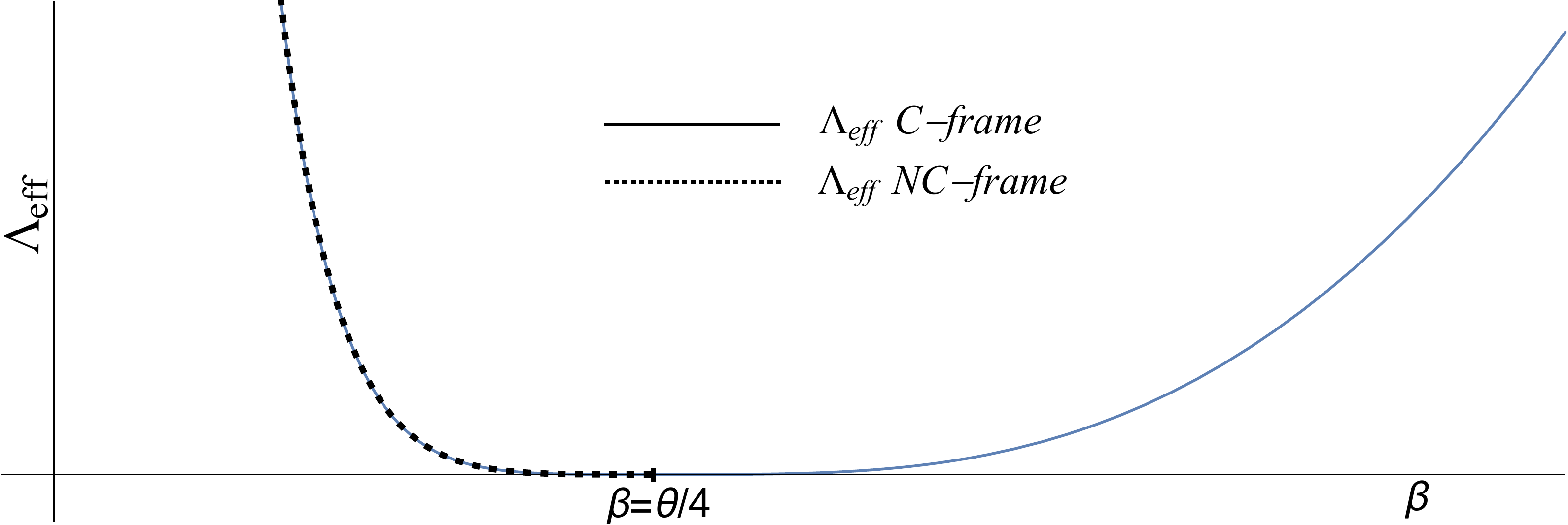}
\end{center}
\caption{\label{fig:de2} 
Plot of $\Lambda_{eff}$ as a function of $\beta$, with fixed values for $\Lambda=1$ and $\theta=0.5$. For large vaues of $\beta$, in the {\it C-frame} $\Lambda_{eff}\sim\beta^2$.}
\end{figure}

 In the {\it C-frame} the volume is always positive, but in order for the volume to be positive in the {\it NC-frame} we need  the condition $\omega'\theta<2$, this further restricts the values of the deformation parameters to $4-\beta\theta>0$. One of the differences in the two frames is the limits in the deformation parameters, therefore the allowed values of $\Lambda_{eff}$ are different.
In the {\it C-frame} for $\beta>\theta/4$ the effective cosmological constant behaves as $\Lambda_{eff}\sim\beta^2$ this can be seen in Fig.(\ref{fig:de2}).

\begin{figure}
\begin{center}
\includegraphics[width=6cm]{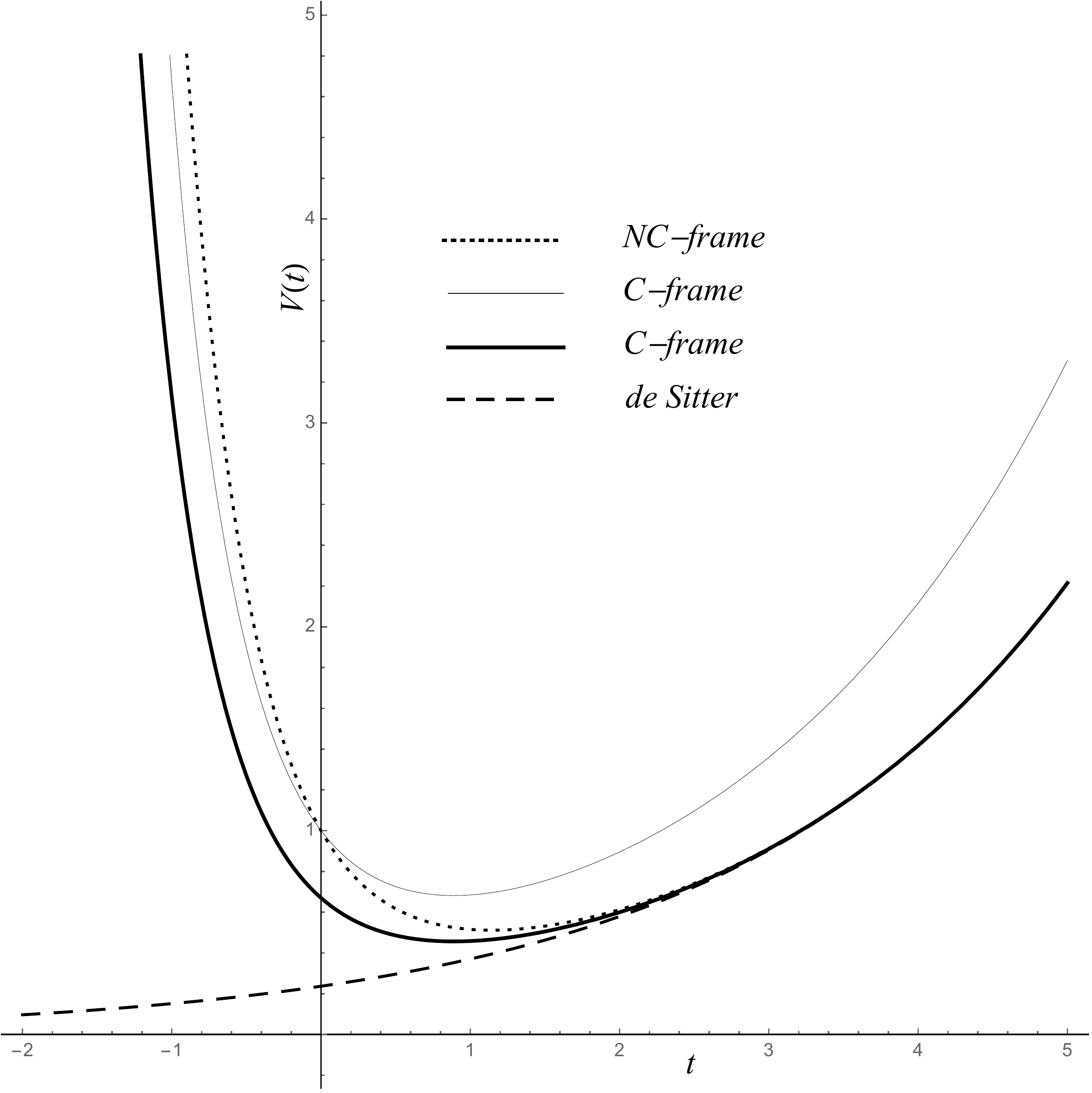}
\end{center}
\caption{\label{fig:de} 
 In all the plots of the deformed model $\kappa=0, \delta_2= \delta_1=0, \beta=1, \Lambda=1$ and $\theta=0.5$.  We can see that for large values of $t$ the behaviour is for the deformed model in the {\it C-frame} the {\it NC-frame} and de Sitter model are the same.}
\end{figure}

To have agreement between the two frames in the asymptotic limit (see Fig.(\ref{fig:de}), we simply redefine the initial volume in the {\it C-frame} as $V_0=\widehat{V}_0(1-\frac{\omega'\theta}{2})$. Therefore  we reach the same physical conclusion in the two frames, that the late time acceleration of the universe is related to the phase space deformations. Therefore in the context of our model, the origin of the cosmological constant is connected to the deformation parameter between the scale factor and the compact dimension. This is analogous  but in a different context, to the results presented in \cite{kallosh}.
Evidence of a possible relationship between the late time acceleration of the universe and the noncommutative parameters has been accumulating \cite{vakili,tachy}, our results also point in this direction. Based on this observation one can argue of a noncommtuative origin for $\Lambda$.
\section{Discussion}

 In this paper we have constructed a deformed phase space model for KK cosmology. The deformation is introduced by modifying the symplectic structure of the minisuperspace variables. This construction is consistent with the assumption taken in noncommutative quantum cosmology \cite{GarciaCompean:2001wy,Barbosa:2004kp,brasil,vakili, tachy}, and enable us to study the effects of phase space deformations in KK cosmology. 

The deformed phase space model is obtained by making the transformation Eq.(\ref{nctrans}) on the canonical Hamiltonian. When writing the deformed Hamiltonian, that depends on the  variables $\widehat{x_i}$ and $\widehat{P_i}$, as a function of the original variables (``{\it C-frame}''), the physical interpretation of the deformation is that of a ghost oscillator in the presence of a``magnetic field" $B$. 
Using the WDW-equation and considering that the frequency of the ghost oscillator is related to the ghost-oscillator frequency (as in \cite{darabi}) and therefore, to an  effective cosmological constant.

Solving the equations of motion for the deformed model and requiring a late time accelerating universe, we find conditions in the parameters $\theta$ and $\beta$. When comparing  the late time behaviour of the ``{\it C-frame}'' and the ``{\it NC-frame}'' descriptions of the deformed model  with the volume of standard de-Sitter cosmology, we get  and effective cosmological constant $\Lambda_{eff}$ as a function of the deformation parameters. It is important to note that this result, which is encoded in Eq.(\ref{lambda}), is the same on the two frames.  For a flat FRW universe,  in the classical as well as the quantum analysis, the effective cosmological constant can behave as $\Lambda_{eff}\sim\beta^2$  and therefore the deformation parameter can play the role of the cosmological constant. Then from the point of view of deformed KK cosmology, the deformation parameters between the scale factor and the compact dimension  can be interpreted as the cosmological constant. Evidence of the origin of $\Lambda$ from deformed phase space cosmology has been accumulating \cite{vakili, tachy} but studies in non cosmological scenarios are needed (i.e. black holes), this question is currently under study and will be reported elsewhere.

\section*{Acknowledgments}
This work is  supported by   CONACYT grants 167335, 179208 and DAIP640/2015. This work is part of the PROMEP research network  ``Gravitaci\'on y F\'isica Matem\'atica".

  \end{document}